\documentclass{eas}
\usepackage{graphics,graphicx}
\TitreGlobal{Submillimetre cosmology}
\newcommand{\arcsecs}{\mbox{$^{\prime\prime}$}}
\newcommand{\arcmin}{\mbox{$^{\prime}$}}
\newcommand{\Lsolar}{\mbox{$L_{\odot}\,$}}
\def\gs{\mathrel{\raise0.35ex\hbox{$\scriptstyle >$}\kern-0.6em\lower0.40ex\hbox{{$\scriptstyle \sim$}}}}
\def\ls{\mathrel{\raise0.35ex\hbox{$\scriptstyle <$}\kern-0.6em\lower0.40ex\hbox{{$\scriptstyle \sim$}}}}

\newcommand{\Msolar}{\mbox{$M_{\odot}\,$}}
\begin{document}

%%-----------------------------
%%      the top matter
%%-----------------------------
\title{Submillimetre cosmology at high angular resolution} 
\author{Thomas R.\ Greve}\address{Max-Planck Institute for Astronomie, K\"onigstuhl 17, D-69117 Heidelberg, Germany}
\begin{abstract}
Over the last decade observations at submillimetre (submm) and millimetre (mm)
wavelengths, with their unique ability to trace molecular gas and dust, have
attained a central role in our exploration of galaxies at all redshifts.  Due
to the limited sensitivities and angular resolutions of current submm/mm
telescopes, however, only the most luminous objects have been uncovered at high
redshifts, with interferometric follow-up observations succeeding in resolving
the dust and gas reservoirs in only a handful of cases.  The coming years will
witness a drastic improvement in the current situation, thanks to the arrival
of a new suite of powerful submm observatories (single-dish and
interferometers) with an order of magnitude improvement in sensitivity and
resolution. In this overview I outline a few of what I expect to be the major
advances in the field of galaxy formation and evolution that these new
ground-breaking facilities will facilitate.
\end{abstract}
\maketitle
%%-----------------------------
%%      your text
%%-----------------------------
\section{Introduction}
Key to understanding the origins of galaxies and their supermassive black holes
(SMBHs) is that benchmark period of cosmic history called the epoch of
reionization (EoR) in which the radiation fields from the first stars and
accreting black holes reionized the surrounding gas -- an epoch of time that
lasted from a couple of hundred million years to about one billion years after
the Big Bang. While the first glimpses into the EoR have now been made owing to
the discovery of a handful of quasars, Lyman-alpha emitters (LAEs) and
gamma-ray bursts (GRBs) at $z\gs 6$ (e.g.\ Fan et al.\ 2003; Kashikawa et al.\
2006; Kawai et al.\ 2006), including the recently discovered GRB at $z\sim 8.3$
(corresponding to about 625 million years after the Big Bang -- Tanvir et al.\
2009), it remains {\it terra incognita} and a top priority of extragalactic
astronomy in the years to come.

Another major challenge for extragalactic astronomy will be to make sense of
the many different types of galaxies found at $z \ls 6 $ in the context of galaxy
evolution after the EoR. It is during this $\sim 12.7\,\rm{Gyr}$ epoch -- i.e.\
from the first one billion years to the present day -- that the bulk growth and
evolution of galaxies and SMBHs took place. High-resolution radio/near-IR/submm
observations have shown that accurate measurements of the sizes, morphologies,
and kinematics of the gas and stars in high-$z$ galaxies are crucial steps in
delineating their evolutionary paths and linking them to galaxy populations at
later cosmic epoch. Yet our understanding of the 'evolutionary tree' of
galaxies is still rudimentary, with even seemingly simple questions such as
'how do galaxies attain their gas?', and 'what is the co-evolution of SMBHs and
galaxies?' lacking definite answers.

While answering these questions will require a 'pan-chromatic' effort, I shall
here focus on the unique contributions we can expect from the next generation
of large submm observatories such as a European 25m aperture submm telescope at
Dome C, the Cornell Caltech Atacama Telescope (CCAT), the Atacama Large
Millimeter Array (ALMA), and the Northern Extended Millimeter Array (NOEMA).
The dramatic improvements in sensitivity, angular resolution, frequency
coverage, and calibration capabilities of these facilities represent a
quantum leap in our ability to map the gas and dust reservoirs in galaxies at
all cosmic epochs, heralding in a new golden era in extragalactic astronomy.

\section{The first galaxies -- peering into the epoch of reionization}
The high cosmic neutral gas fraction ($f_{\mbox{\tiny{HI}}}\gs 10^{-3}$) at
$z\gs 6$ -- inferred from the observed Gunn-Peterson absorption troughs towards
high-$z$ QSOs (e.g.\ Fan et al.\ 2003) -- implies that only observations longward
of $\sim 0.9\,\mu\rm{m}$ (and X-rays) are able to peer into the EoR. 

Continuum observations probing the thermal dust emission at submm/mm
wavelengths are particular powerful for this endeavour, given their unique
negative $k$-correction which effectively offsets the cosmic dimming out to
redshifts of $z\sim 10$ (Blain \& Longair 1993).  The submm detection of the $z
= 6.42$ QSO SDSS\,J1148$+$5251 (Walter et al.\ 2004) demonstrated the presence
of a large amount of dust ($\sim 10^9\,\Msolar$), pointing to rapid dust
production mechanisms at these early epochs.  Type II supernovae (SNe) has been
suggested as an important source of dust (Dunne et al.\ 2003), although
according to models the contribution from AGB stars may also be substantial
(Valiante et al.\ 2009).  ALMA will probe $\gs 10\times$ (Fig.\ 1) further down
the IR luminosity function at these redshifts than what is currently possible,
detecting the rest-frame FIR/submm emission from 'normal' low-luminosity ($\sim
10^{10}\,\Lsolar$) galaxies, e.g.\ LAEs which models predict will be
sufficiently submm bright ($S_{\mbox{\tiny{850$\mu$m}}}\sim 0.1\,\rm{mJy}$ --
Dayal et al.\ 2009).  The superior continuum sensitivity of ALMA over a broad
mm/submm range ($84-950\,\rm{GHz}$) combined with its remarkable angular
resolution ($\sim 0.1\arcsecs$, corresponding to $\sim 0.6\,\rm{kpc}$ at $z
\sim 6$) will allow us to spatially resolve the dust emission, disentangle
hot/cold dust components, determine accurate dust masses for large samples of
$z > 6$ galaxies, and ultimately reveal the dust production processes and
time-scales in the early Universe.
\begin{figure}
\begin{center}
\includegraphics{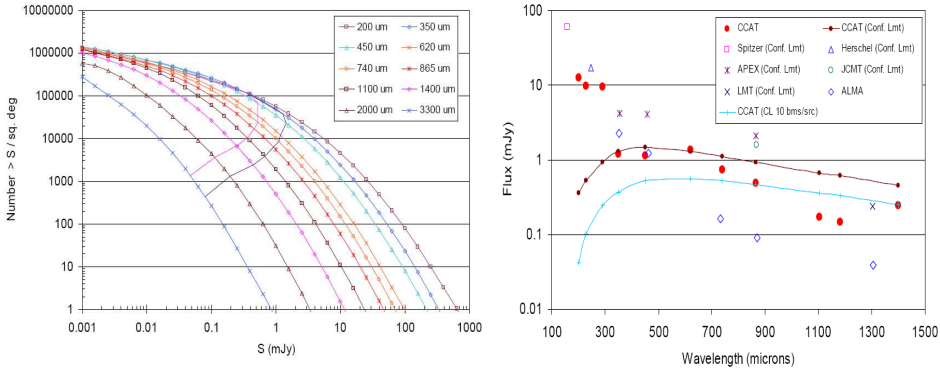}
\end{center}
\vspace*{4.2cm}
\caption{{\bf Left:} Models of the submm number counts at different wavelengths
(Blain et al.\ 1999, 2002). The crossing lines show 30 (lower) and 10 (upper)
beams/source confusion limits for a 25m single-dish telescope. {\bf Right:}
Continuum point source sensitivities (5-$\sigma$) after 1hr integration vs.\
wavelength for CCAT (red circles) and ALMA (blue diamonds), along with
confusion limits (30 beams/source) for a number of other submm/mm facilities.
Adapted from the CCAT Science Case (v1.3).  
}
\label{figure:Fig1}
\end{figure}

Turning to the prospects of spectral line studies of galaxies within the EoR:
the detections of low- to mid-$J$ CO lines in a handful of $z\sim 4-6$ QSOs by
Wei\ss~et al.\ (2006) led them to predict (based on extrapolation of models)
that a) the $J\gs 7$ lines are not significantly excited, and b) if similar
excitation conditions prevail in $z\gs 7$ objects, their CO detection-potential
with ALMA is slim (due to its low-frequency cut-off at $84\,\rm{GHz}$, which
means only $J > 7$ lines are observable at $z > 7$).
In the local Universe the nuclei of starburst and seyfert galaxies are found to
have much higher CO line ratios (i.e.\ excitation conditions) than the global
galaxy average (e.g.\ Papadopoulos \& Seaquist 2000). Since few of the high-$z$
QSOs have been robustly resolved in CO, it is possible that the same 'aperture'
effects are at play here (and affecting the modeling), with the high-$J$ lines
being excited in the nuclear/starburst regions only. The sensitivity and
resolution of ALMA (and NOEMA) should be able to spatially isolate this nuclear
high-$J$ CO emission from the galaxy regions further out, allowing for a 
more accurate picture of the ISM conditions in the first QSOs.

While QSOs are likely to harbor the most extreme excitation conditions
possible, more normal high-$z$ galaxies are likely to have quiescent, possibly
Milky Way-type, ISM conditions. Due to selection effects (bias towards
luminous, mid- to high-$J$ CO lines at high $z$), only a few examples of such
galaxies have been found at $z > 1$ (Papadopoulos \& Ivison 2002; Greve et al.\
2003; Dannerbauer et al.\ 2009), and as first pointed out by Papadopoulos \&
Ivison (2002), galaxies with such ISM conditions at $z\gs 7$ would be extremely
difficult to detect in CO with ALMA. The recent detections of the [C{\sc ii}]
158\,$\mu$m fine-structure line towards a handful of $z > 4$ QSO (Maiolino et al.\ 2005, 2009;
Iono et al.\ 2006; Walter et al.\ 2009) have raised hopes for a more feasible
way to study $z\gs 7$ objects in the future.  Not only is the [C{\sc ii}] line
observable with ALMA at these redshifts, but in SDSS\,J1148$+$5251 the line is
$\sim 5\times$ brighter than the strongest CO line ($J=6-5$).  Furthermore,
[C{\sc ii}] not only remains luminous in metal-poor systems (where CO is
photo-dissociated by UV photons), which would favor its detection in
low-mass systems, but it is also not exclusively tied to star-forming sites; in
fact the cold and warm neutral medium as well as ionized gas (H\,{\sc ii}
regions) can have significant [C{\sc ii}] emission contributions (Madden et
al.\ 1997), in which case sensitive, high-resolution observations could reveal
'hidden' reservoirs of gas that CO would not pick up.

\section{Galaxy evolution after the first one billion years}
\subsection{The dust-obscured Universe}
The first extragalactic submm (850-$\mu$m) surveys with SCUBA/JCMT (Barger et
al.\ 1998; Hughes et al.\ 1998) showed that a significant fraction of the
cosmic star formation at $z\sim 1-3$ occurred in heavily dust-obscured
galaxies, so-called submm-selected galaxies (SMGs -- Blain et al.\ 2002), that
had been missed by UV/optical surveys. Such surveys, with their coarse
resolution ($\rm{FWHM}\sim 15\arcsecs$) and confusion limit ($\sim 2\,$mJy),
only select the most luminous ($\sim 10^{13}\,\Lsolar$) objects of the
population, although surveys using galaxy clusters as 'cosmic magnification
glasses' have been able go somewhat beyond the formal confusion limit of
blank-field surveys (reaching depths of $\sim 1\,$mJy at 850-$\mu$m -- Smail et
al.\ 2002; Chapman et al.\ 2003; Knudsen et al.\ 2008).

However, the next generation of single-dish submm telescopes -- i.e.\ 20-30\,m
single-dish antennas located at Chajnantor and/or Dome C -- equipped with
large-format filled-array cameras capable of effectively operating at $\sim
200-850\,\mu$m would have the sufficient angular resolution ($\sim
2-4\arcsecs$) and field of view ($\sim 10\arcmin-20\arcmin$) to map several
square degrees of sky down to depths of $\ls 0.5\,\rm{mJy}$ (Fig.\ 1), thereby
a) uncover tens of thousands of sub-mJy SMGs, b) sample the entire 'dust'
luminosity function, and c) resolve the entire FIR/submm background.  The
relatively small beam sizes of such large single-dish telescopes, combined with
observations in multiple submm bands would allow for robust counterpart
identifications and photometric redshift estimates (Hughes et al.\ 2002;
Aretxaga et al.\ 2003).  Interferometers with their small fields of view are
unlikely useful blank-field survey facilities. For example, given the primary
beam of ALMA ($\sim 9\arcsecs - 56\arcsecs$ from Band 9 to 3), a single, deep 
($\sigma \sim 3\times 10\,\mu\rm{Jy}$) 'blank' pointing is expected to yield no
more than 1-2 sources -- making it very time-consuming to robustly constrain
the faint number counts.  Instead, the main strength of submm/mm
interferometers will be as follow-up facilities, providing extremely
high-resolution and high-fidelity continuum imaging of the dust reservoirs in
pre-selected galaxies.

The FIR/submm spectral energy distributions (SEDs) of high-$z$ galaxies are
usually inferred from fits of simple, optically thin dust emission models to a
few ($ < 4$) {\it global}, i.e.\ unresolved, photometric data points.  However,
it has been suggested that the dust in local Ultra Luminous Infra-Red Galaxies
($L_{\mbox{\tiny{IR}}}\ge 10^{12}\,\Lsolar$ -- ULIRGs) could be optically thick
at FIR/submm wavelengths (Condon et al.\ 1991; Solomon et al.\ 1997; Lisenfeld
et al.\ 2003), in which case a revision of of dust mass and temperature
estimates may be in store, as would the usefulness of FIR/submm line
diagnostics. In SMGs where the gas and dust may be even more densely packed
than in local examples (cf.\ Iono et al.\ 2008), optical depth effects may play
an even bigger role.  The problem of dust opacity will be addressed by future
submm interferometers, by resolving the dust emission on sub-kpc scales within
galaxies over a broad range of frequencies (Fig.\ 2). Furthermore, such dust
maps would match the resolution of current radio observations, allowing for
resolved FIR-radio correlation (Condon 1992) studies in galaxies out to $z\sim
3$ (the limit is set by the radio, not the submm, observations). This would
shed important new light on the underlying physics of the correlation, e.g.\
its dependence on the the initial mass function and the magnetic field vs.\ CMB
energy density ratio.

\subsection{Galaxy formation:  major mergers or cold accretion?}
How did galaxies accrete their gas -- was it primarily via one or two major
mergers, interspersed with periods of quiescent evolution, or did they assemble
and evolve through a prolonged, steady 'cold' accretion of gas? Observational
evidence for both scenarios is found at high redshifts, yet the details of
these two modes of galaxy formation remain unclear, as does their
relative importance with redshift, galaxy/halo mass, and environment.\\

\noindent There is now strong observational evidence in favour of bright
($S_{\mbox{\tiny{850$\mu$m}}}\gs 5\,\rm{mJy}$) SMGs being the progenitors of
massive spheroids. Their extreme star formation rates ($\sim
1000\,\Msolar\,\rm{yr}^{-1}$) and their co-moving number density ($\sim
10^{-5}\,\rm{Mpc}^{-3}$), which is similar to that of giant ellipticals at
later epochs, both point in this direction. Moreover, interferometric CO
observations have determined gas masses ($M_{gas} \sim 3\times 10^9\,\Msolar$),
physical sizes ($R_{1/2} \sim 2\,$kpc), line widths ($\rm{FWHM}\sim
700\,\rm{km}\,\rm{s}^{-1}$), dynamical masses ($M_{dyn}\sim 10^{11}\,\Msolar$
within $R_{1/2}$) and gas fractions ($f_{gas}\sim 0.4$), all of which strongly
suggest that SMGs are gas-rich, highly-dissipative major mergers in which the
gas is rapidly funneled towards the center due to angular momentum loss via
dissipation, igniting an intense 'maximal' starburst capable of producing a
$2-3 m_*$ galaxy within a few hundred million years (Neri et al.\ 2003; Greve
et al.\ 2005; Tacconi et al.\ 2006, 2008; Swinbank et al.\ 2006).  The inferred
central surface densities in SMGs, however, are $\sim 10\times$ those observed
in $z\sim 0$ spheroids (Bouch\'e et al.\ 2007; Tacconi et al.\ 2008), which
have led to the suggestion that $z\sim 2$ SMGs are the direct progenitors of
the recently discovered population of extremely compact ($R\sim 1\,\rm{kpc}$),
very dense, 'dead' galaxies at $z\sim 1.5$ (Trujillo et al.\ 2006; Zirm et al.\
2007; Toft et al.\ 2007). The latter,  in order to reach the larger half-light
sizes and lower central densities of today's early-type galaxies, will have had
to undergo 'dry' accretion, i.e.\ mergers where little or no gas is involved
(van Dokkum 2005; Naab et al.\ 2007), or a combination of both (Burkert et al.\
2008). The above evolutionary scenario is still very tentative, however, and in
fact we note that far from all SMGs are compact, some have CO and radio
morphologies extended up to $\sim 5-10\,\rm{kpc}$ scales (Chapman et al.\ 2004;
Bothwell et al.\ in prep.; Fig.\ 2).

Major mergers also seem to be associated with the most luminous high-$z$ radio
galaxies (HzRGs) and optically selected QSOs, a third of which have been shown
to be extremely IR-luminous ($L_{\mbox{\tiny{IR}}}\simeq 10^{13}\,\Lsolar$ --
Omont et al.\ 2003).  Subsequent high-resolution interferometric submm
observations of a subset of those have found massive ($\sim 10^{10}\,\Msolar$),
but often spatially and kinematically distinct, gas and dust components
(Papadopoulos et al.\ 2000; Riechers et al.\ 2008). While this is indicative of
an ongoing major 'wet' merger, the overall masses and luminosities involved are
substantially larger than those of SMGs, giving the impression that we are
dealing with a much more violent event, possibly a 'monolithic' collapse of a
giant gas cloud which occurs in only the most highly mass-biased and rare peaks
of the cosmic density field.\\
\begin{figure}
\begin{center}
\includegraphics{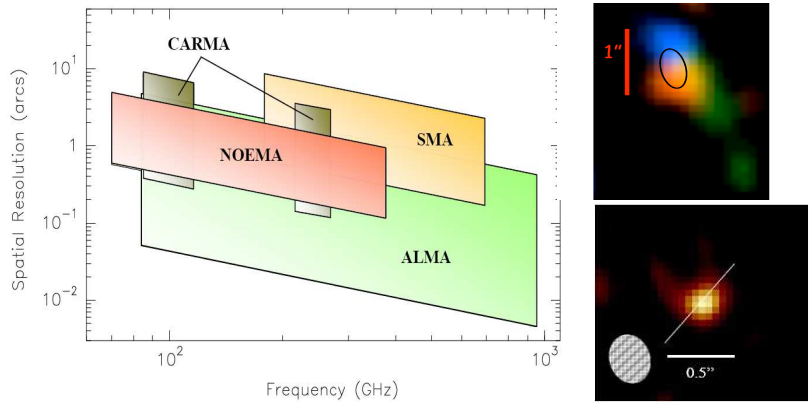}
\end{center}
\vspace*{4.7cm}
\caption{{\bf Left:} Angular resolution vs.\ observing frequency for current
submm/mm interferometers (SMA, CARMA) and future facilities (NOEMA, ALMA).
Adopted from the NOEMA Phase-A study document.  {\bf Right:} High-resolution CO
(3--2) (top) and (6--5) (bottom) maps of SMGs at $z=1.53$ (Bothwell in prep.) and $z=2.20$ 
(Tacconi et al.\ 2008), respectively. The CO (3--2) map is color coded by velocity. 
}
\label{figure:Fig2}
\end{figure}

\noindent Observational evidence for galaxy growth via the steady accretion of
gas flows and minor mergers has come from kinematically resolved H$\alpha$
surveys of UV/optical selected galaxies at $z\sim 1-3$ using integral field
units (e.g.\ F\"{o}rster-Schreiber et al.\ 2006; Genzel et al.\ 2006).  While
these galaxies are as massive as SMGs ($M_{dyn}\sim 10^{11}\,\Msolar$), their
morphologies and kinematics differ fundamentally from SMGs in that they have
disk-like shapes and show (H$\alpha$) velocity fields resembling that of disks,
with some being dominated by rotation (circular velocities of $v_c \sim
250\,\rm{km}\,\rm{s}^{-1}$) and others by turbulent motion (velocity
dispersions of $\sigma \sim 50\,\rm{km}\,\rm{s}^{-1}$).  This picture is
supported by recent interferometric CO observations towards a few of these
galaxies (Daddi et al.\ 2008; Dannerbauer et al.\ 2009), which found
(tentative) evidence for extended emission ($R_{1/2}\sim 4\,\rm{kpc}$) and
rotation consistent with a disk. In addition, evidence of low-excitation gas
(subthermal CO $3-2/1-0$ line ratio) was found, akin to the 'moderate' ISM
conditions in local spirals and our own Galaxy.  Furthermore, their average
star formation efficiencies (as gauged by
$L_{\mbox{\tiny{IR}}}/L'_{\mbox{\tiny{CO}}}$) was found to be $4-5\times$ below
that of SMGs, but consistent with that of local spirals.  Simulations suggest
that the high degree of gas turbulence found in some of these high-$z$ disks
($v_c/\sigma \ls 3$, compared to $v_c/\sigma \sim 10$ in local disks) may be
driven by the kinetic energy gained by the gas as it streams inwards along
filaments, in turn causing disk-instability and the subsequent break-up into
giant clumps of gas, capable of forming stars efficiently (Dekel et al.\
2009).\\ 

\noindent With ALMA and NOEMA it will be possible to very effectively target
large samples of pre-selected high-$z$ galaxies across a broad range of
redshifts, masses and selection criteria, and by means of a rich variety of ISM
diagnostic molecular (e.g.\ CO, HCN, HCO$^+$) and atomic lines (e.g.\ [C\,{\sc
ii}] 158\,$\mu$m and [O\,{\sc i}] 63 and 145\,$\mu$m), accurately measure their
(resolved) velocity fields, dynamical masses, gas fractions and morphologies.
The deep understanding of the gas accretion processes in high-$z$ galaxies, and
their evolutionary links to intermediate- and low-$z$ galaxy populations, is
just one major result that will be gleaned from such studies. They will also
provide a rather exhaustive account of the physical properties of the ISM
(e.g.\ density and temperature regimes, heating and cooling mechanisms, UV
field strength). Linking these to the distributed star formation (traced by
dust emission) we can explore changes in the slope of the star formation law
for different types of high-$z$ galaxies and different gas density tracers, and
whether the slope goes from superlinear to linear once sub-kpc scales are
probed (Bigiel et al.\ 2008).

\section{Supermassive black holes and their host galaxies}
\subsection{The origin of supermassive black holes} 
Most galaxies in the present-day Universe harbour a SMBH in their centers
(e.g.\ Kormendy et al.\ 1996, 1997), yet the origins of these black holes
remain a mystery. In all likelihood, however, they grew from 'seed' black holes
that formed very early in the Universe's history ($z > 10$). Intermediate-mass
($\sim 10^2\,\Msolar$) BHs resulting from Pop III stars going supernova, may
have been such 'seeds' for merger trees to SMBHs (e.g.\ Ohkubo et al.\ 2009).
In an alternative scenario the monolithic gravitational collapse of primordial
gas clouds led to the formation of 'seeds' with initial masses of $\sim
10^5\,\Msolar$ (Bromm \& Loeb 2003). The latter scenario might be favored as
the progenitors of the $\sim 10^9\,\Msolar$ SMBHs found in $z\gs 6$ QSOs (e.g.\
Becker et al.\ 2001), as it is difficult to make Pop III BHs grow so massive in
only a few hundred million years even if they are accreting at super-Eddington
rates. One may speculate, however, whether Pop III remnants provided the
'seeds' for the SMBHs in less massive, but more abundant, galaxies, and whether
the SMBH growth in these galaxies occurred at a much more leisurely rate?

It was recently proposed that the high-$J$ rotational transitions of CO, HCN,
HCO$^+$ as well as the [C\,{\sc ii}] and [O\,{\sc i}] fine-structure lines, may
be used as markers of distant SMBHs (Meijerink \& Spaans 2005, 2006).  As a
general result -- valid over a broad swath of gas conditions -- models predict
that the high-$J$ ($J > 6$) lines are much more excited in X-ray dominated
regions (XDRs -- Maloney et al.\ 1996) where the gas is exposed to an intense,
hard X-ray radiation field ($E = 1-100\,$keV) characteristic of an AGN, than in
environments  where the gas is being irradiated by FUV-photons ($E =
6-13.6\,$keV) representative of the Photo-dominated regions (PDRs -- Tielens \&
Hollenbach 1985) found in OB star forming regions.  The reason for this is the
higher gas temperatures ($10^2-10^3\,$K) in XDRs caused by fast, penetrative
electrons produced via X-ray photo-ionization, compared to the much more modest
heating in PDRs via photo-electric emission from dust grains and UV pumping of
H$_2$. 

While this is a novel and untested technique, it is worth noticing that
interferometric studies of CO, HCN and HCO$^+$ emission towards a small number
of local IR-luminous galaxies where the presence of a powerful AGN is known by
some other means, found elevated HCN and HCO$^+$ 1--0 line fluxes (Kohno et
al.\ 2003; Imanishi et al.\ 2007), in qualitative agreement with the enhanced
HCN abundance expected in a XDR-type chemistry (Lepp \& Dalgarno 1988). The
high-$J$ lines, however, which are expected to be even stronger AGN vs.\
starburst discriminants, have so far not been investigated in local galaxies
due to opaqueness of the atmosphere at $\gs 400$\,GHz.  ALMA will have the
frequency coverage and sensitivity to detect the high-$J$ lines towards
galaxies out to $z \gs 10$ (Spaans \& Meijerink 2008). High angular resolution
($\ls 0.1\arcsecs$, corresponding to $\ls 500\,\rm{pc}$ at $z\sim 10$) will be
important for such AGN vs.\ starburst studies, since the high-$J$ CO emitting
gas associated with a XDR/AGN is likely to be compact ($R \ls 500\,$pc), and
blindly modeling the global line ratios is therefore prone to a 'contaminating'
PDR contribution from distributed starforming gas.

\subsection{The early co-evolution of SMBHs and their host galaxies}
Optical spectroscopy of spheroids in the local Universe have established tight
relations between SMBHs and the luminosities, masses and velocity dispersions
of their host galaxies (Magorrian et al.\ 1998; Gebhardt et al.\ 2000),
suggestive of a self-regulating, physical connection between SMBH growth and
spheroid build-up.  High-resolution maps of the gas morphology and kinematics
in the central regions of high-$z$ galaxies will allow us to study AGN fueling
mechanisms and feedback processes on sub-kpc scales, thereby playing an important
role in probing the physics behind the above scaling relations and their
evolution with redshift.

Optical/NIR measurements of spheroid masses ($M_{\mbox{\tiny{sph}}}$) at
high-$z$ are difficult to come by, since the luminous AGN tends to 'drown out'
any other light. Instead, some of the strongest
$M_{\mbox{\tiny{sph}}}$-constraints have come from dynamical mass estimates
derived from high-resolution CO (or [C{\sc ii}]) maps of IR-luminous $z > 4$ QSOs (Walter et
al.\ 2003, 2009; Riechers et al.\ 2008), leading to the first indications that
luminous QSOs at these early epochs had
$M_{\mbox{\tiny{BH}}}/M_{\mbox{\tiny{sph}}}$ ratios $\gs 10\times$ above the
local relation ($M_{\mbox{\tiny{BH}}}/M_{\mbox{\tiny{sph}}}\simeq 1.4\times
10^{-3}$). CO observations of optically luminous ($M_{\mbox{\tiny{B}}}\sim
-28$), submm-detected QSOs at $z\sim 2$ find
$M_{\mbox{\tiny{BH}}}/M_{\mbox{\tiny{sph}}}\sim 9\times 10^{-3}$ (Coppin et
al.\ 2008), i.e.\ also significantly above the local ratio.  The CO
gas-dynamics of submm-detected, but optically less luminous
($M_{\mbox{\tiny{B}}}\sim 25$) QSOs at $z\sim 2$, however, suggest that they
lie within the local relation given the uncertainties.  Finally,
high-resolution CO studies of SMGs (in conjunction with X-ray and optical/NIR
observations) have shown that SMGs lie $\sim 3-5\times$ below the local
$M_{\mbox{\tiny{BH}}}-M_{\mbox{\tiny{sph}}}$ relation (Alexander et al.\ 2008).
Thus, it appears that the $M_{\mbox{\tiny{BH}}} - M_{\mbox{\tiny{sph}}}$
relation evolves with redshift, and that it does so differently for different
types of galaxies. In optically luminous QSOs, the build-up of the spheroid
stellar mass lags the black hole growth, while in SMGs the situation appears to
be reversed.  
\begin{figure}[t]
\begin{center}
\includegraphics{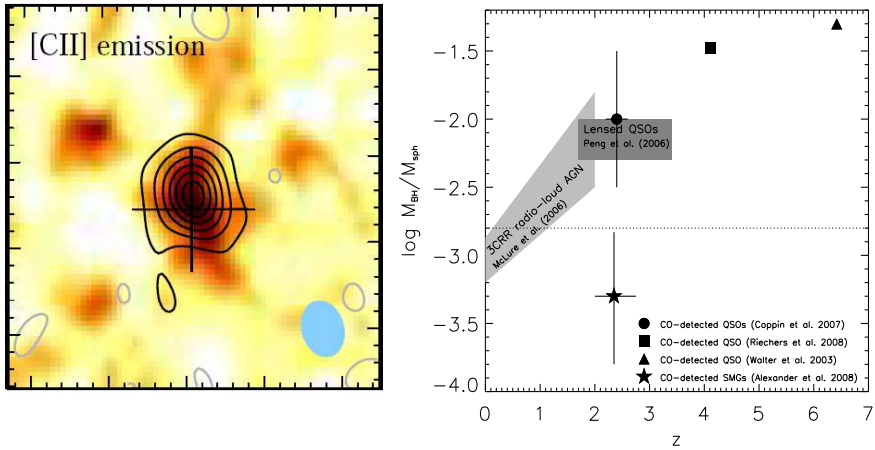}
\end{center}
\label{figure:Fig3}
\vspace*{5.7cm}
\caption{{\bf Left:} High-resolution ($\sim 0.3\arcsecs$ FWHM) maps of the
[C{\sc ii}] (contours) and CO(3--2) (color scale) emission toward the $z=6.42$ QSO
J1148. Adapted from Walter et al.\ (2009).  {\bf Right:} Observational
constraints from high-resolution CO studies on the the SMBH-spheroid mass ratio
for different types of galaxies at various redshifts. The dotted line
represents the local ratio and no evolution.  Adapted from Coppin et al.\
(2008). 
}
\end{figure}

Substantial uncertainties are associated with the above findings, however,
owing to the often poorly constrained CO (or [C\,{\sc ii}]) sizes and
inclination angles.  More worrying, though, for such potentially dynamically
unsettled (non-virialized) systems, is the possibility that the gas
might not always be centered around the black hole or even be a good probe of
the dynamical mass. Examples of such AGN/H$_2$ configurations have recently
been found at both high and low $z$ (Ivison et al.\ 2008; Riechers et al.\
2008; Papadopoulos et al.\ 2008). ALMA, with its ability to spatially and
kinematically resolve the gas reservoirs in distant galaxies on sub-kpc scales
at high $S/N$, will explore these effects in great detail, allowing for more
accurate spheroid and dynamical mass estimates in less luminous systems than is
currently possible, thus facilitating a more straightforward comparison with
local galaxies.   

Estimating SMBH masses will remain challenging, however, and will initially at
least be limited to QSOs/type-I AGN and SMGs where locally calibrated
techniques for measuring black hole masses, such as reverberation mapping
(Onken et al.\ 2004; Greene \& Ho 2005) and single epoch virial estimates
(Vestergaard \& Peterson 2006), can be applied.  With the advent of James Webb
Space Telescope (JWST), however, there is hope that significantly improved SMBH
mass estimates will be possible from direct imaging of the (atomic) gas and
stellar kinematics (Ferrarese \& Ford 2005).  Studying the
(mass-)interrelationship of SMBHs and their host galaxies as a function of
redshift and luminosity is likely to become an area of great synergy between
ALMA and JWST.  By contrasting the various SMBH-spheroid correlations to model
predictions of their evolution (Di Matteo et al.\ 2005; Granato et al.\ 2004;
Dotti et al.\ 2007) we will gain a much deeper insight into the underlying
physics and the role SMBH and AGN-feedback plays in the overall scheme of
galaxy formation and evolution.

\acknowledgements
{\bf Acknowledgments} I'm grateful to the organizers and participants of the
ARENA meeting for an interesting and stimulating meeting. I thank Matt Bothwell
and Scott Chapman for providing me with a figure from Bothwell et al.\ in prep.
I'm also thankful to F.\ Walter, C.\ Carilli, R.\ J.\ Ivison, P.\ P.\
Papadopoulos for useful comments.

%%-----------------------------
%%      your bibliography
%%-----------------------------


\begin{thebibliography}{99}
\bibitem[2002]{Blain-et-al-2002} Blain, A.W.\ \etal\ 2002, PhR, 369, 111
\bibitem[2009]{Bouche-et-al-2009} Bouch\'e, N.\ \etal\ 2009, ApJ, 671, 303
\bibitem[2004]{Chapman-et-al-2004} Chapman, S.C.\ \etal\ 2004, ApJ, 614, 671
\bibitem[2008]{Coppin-et-al-2008} Coppin, K.E.K.\ \etal\ 2008, MNRAS, 389, 45
\bibitem[2008]{Daddi-et-al-2008} Daddi, E.\ \etal\ 2008, ApJ, 673, L21
\bibitem[2003]{Fan-et-al-2003} Fan, X.\ \etal\  2003, AJ, 125, 1649
\bibitem[2006]{Genzel-et-al-2006} Genzel, R.\ \etal\  2006, Nature, 442, 786
\bibitem[2008]{Ivison-et-al-2008} Ivison, R.J.\ \etal\  2008, MNRAS, 390, 1117
\bibitem[2006]{Kashikawa-et-al-2006} Kashikawa, N.\ \etal\  2006, ApJ, 468, 7
\bibitem[2006]{Kawai-et-al-2006} Kawai, N.\ \etal\  2006, Nature, 440, 184
\bibitem[1996]{Kormendy-et-al-1996} Kormendy, J.\ \etal\  1996, ApJ, 459, L57
\bibitem[2009]{Maiolino-et-al-2009} Maiolino, R.\ \etal\ 2009, A\&A, 500, L1
\bibitem[2005]{Meijerink-and-Spaans-2005} Meijerink, R.\ \& Spaans, M.\ 2005, A\&A, 436, 397
\bibitem[2007]{Naab-et-al-2007} Naab, T.\ \etal\  2007, ApJ, 658, 710
\bibitem[2003]{Neri-et-al-2003} Neri, R.\ \etal\ 2003, ApJ, 597, L113
\bibitem[2008]{Riechers-et-al-2008} Riechers, D.A.\ \etal\ 2008, ApJ, 686, L9
\bibitem[1997]{Solomon-et-al-1997} Solomon, P.M.\ \etal\ 1997, ApJ, 478, 144
\bibitem[2002]{Papadopoulos-and-Ivison-2002} Papadopoulos, P.P.\ \& Ivison, R.J.\ 2002, ApJ, 564, L9
\bibitem[2006]{Swinbank-et-al-2006} Swinbank, A.M.\ \etal\ 2006, MNRAS, 371, 465
\bibitem[2002]{Smail-et-al-2002} Smail, I.\ \etal\ 2002, MNRAS, 331, 495
\bibitem[2008]{Tacconi-et-al-2008} Tacconi, L.J.\ \etal\  2008, ApJ, 680, 246
\bibitem[2009]{Tanvir-et-al-2009} Tanvir, N.R.\ \etal\  2009, Nature, 461, 1254
\bibitem[2007]{Toft-et-al-2007} Toft, S.\ \etal\  2007, ApJ, 671, 285
\bibitem[2006]{Trujillo-et-al-2006} Trujillo, I.\ \etal\  2006, MNRAS, 373, L36
\bibitem[2009]{Walter-et-al-2009} Walter, F.\ \etal\  2009, Nature, 457, 699
\end{thebibliography}
\end{document}